\documentclass[runningheads]{llncs}
\usepackage{graphicx}
\begin{document}
\title{Improved automated lesion segmentation in whole-body FDG/PET-CT via Test-Time Augmentation}
%
%
\author{Sepideh Amiri\inst{1}\orcidID{0000-0002-2848-4671} \and Bulat Ibragimov\inst{1}\orcidID{0000-0001-7739-7788}}
\authorrunning{Amiri et al.}
%
\institute{Department of Computer Science, University of Copenhagen, Copenhagen 2100, Denmark \\
\email{\{sepideh.amiri,bulat\}@di.ku.dk}}
\maketitle              
\begin{abstract}
Numerous oncology indications have extensively quantified metabolically active tumors using positron emission tomography (PET) and computed tomography (CT). F-fluorodeoxyglucose-positron emission tomography (FDG-PET) is frequently utilized in clinical practice and clinical drug research to detect and measure metabolically active malignancies. The assessment of tumor burden using manual or computer-assisted tumor segmentation in FDG-PET images is widespread. Deep learning algorithms have also produced effective solutions in this area. However, there may be a need to improve the performance of a pre-trained deep learning network without the opportunity to modify this network. We investigate the potential benefits of test-time augmentation for segmenting tumors from PET-CT pairings. We applied a new framework of multilevel and multimodal tumor segmentation techniques that can simultaneously consider PET and CT data. In this study, we improve the network using a learnable composition of test time augmentations. We trained U-Net and Swin U-Netr on the training database to determine how different test time augmentation improved segmentation performance. We also developed an algorithm that finds an optimal test time augmentation contribution coefficient set. Using the newly trained U-Net and Swin U-Netr results, we defined an optimal set of coefficients for test-time augmentation and utilized them in combination with a pre-trained fixed nnU-Net. The ultimate idea is to improve performance at the time of testing when the model is fixed. Averaging the predictions with varying ratios on the augmented data can improve prediction accuracy. Our code will be available at \url{https://github.com/sepidehamiri/pet\_seg\_unet}

\end{abstract}

\section{Introduction}
For computer-assisted cancer detection and treatment, automatic tumor segmentation from medical images is a crucial step. Deep learning has recently been effectively used for this problem, improving performance~\cite{jiang2021deep}. However, most deep learning segmentation techniques currently in use are limited to one imaging modality. Today's clinics frequently use PET/CT scanners, which combine PET and CT into one device and deliver metabolic and anatomical data. The particular challenge of lesion segmentation in FDG-PET resides in the fact that healthy organs, such as the brain, bladder, etc, can have high FDG uptake, making it challenging to avoid false positive segmentations, which can be seen in Fig~\ref{fig1}. Various studies have been proposed to segment tumors in PET/CT scans autonomously. To get constant segmentation masks between PET and CT, Song et al. created an adaptive context term for the target function~\cite{song2013optimal}. In order to get object seeds, Ju et al. adopted a random walk approach as an initial preprocessing. After that, a graph cut method was applied to segment lung tumors on PET/CT images~\cite{ju2015random}. Based on the Markov Random Field optimization issue, Han et al. developed a PET/CT segmentation formulation~\cite{han2011globally}. All of the aforementioned studies showed that integrating the data from multiple imaging modalities might produce tumor segmentation results that are more precise than the segmentation results obtained from a single image modality.

\section{Methods and Materials}
\subsection{Dataset}
\begin{figure}
\begin{center}
\includegraphics[width=0.6\textwidth]{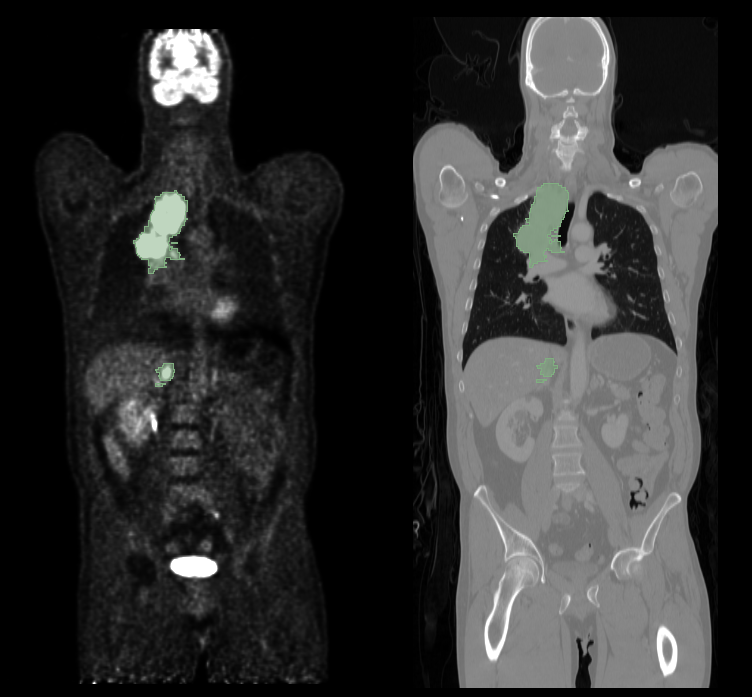}
\caption{An illustration of fused whole-body FDG-PET/CT data. The manually segmented malignant lesions are shown in the green sections.} \label{fig1}
\end{center}
\end{figure}

We used an annotated oncologic PET/CT data set in this study. Between 2014 and 2018 at the University Hospital Tübingen, 501 consecutive whole-body FDG-PET/CT data sets of patients with malignant lymphoma, melanoma, and non-small cell lung cancer (NSCLC) and 513 data sets without PET-positive malignant lesions (negative controls) were studied~\cite{gatidis2022whole}. Additionally, 60 minutes after receiving an I.V. injection of 300–350 MBq 18F-FDG, a full-body FDG-PET scan was performed for each patient. PET data were rebuilt using the ordered-subset expectation maximization (OSEM) technique, which had a gaussian kernel of 2 mm, 21 subsets, and two iterations on a 400 x 400 matrix. Fig.~\ref{fig1} shows an example of fused whole-body FDG-PET/CT data.
\subsection{Preprocessing}
The used dataset already had pre-processing, including normalized and resampled (CT to PET imaging resolution; same matrix size) (PET converted to standardized update values; SUV). By converting image units from activity counts to standardized uptake values, PET data were made uniform (SUV). We also applied intensity scaling with a minimum of 100 and 0 and a maximum of 250 and 15 for CT and PET, respectively. To reduce the model's memory and computation requirements during segmentation, we cropped the images' foreground based on the CT.

\subsection{Network Architecture}
\begin{figure}
\begin{center}
\includegraphics[width=1\textwidth]{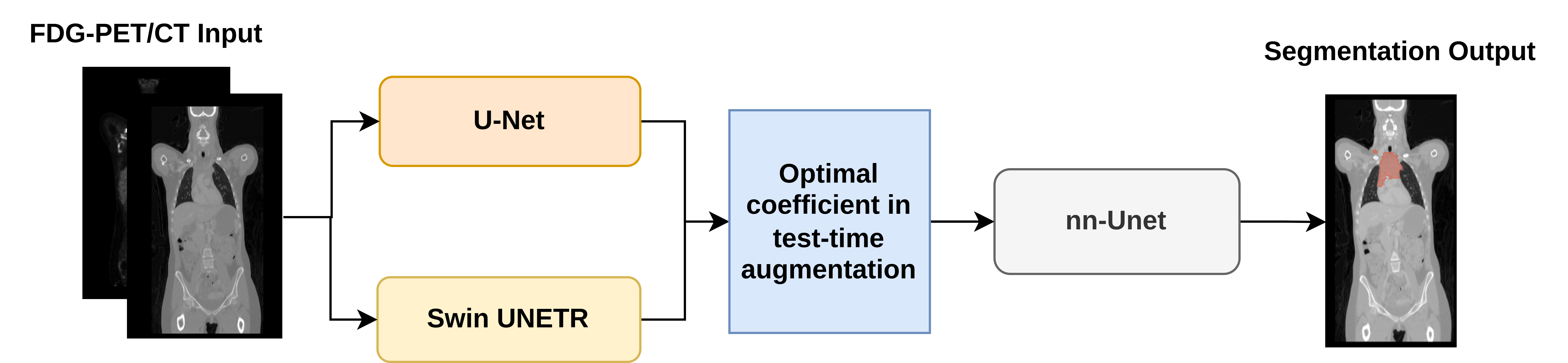}
\caption{Our framework trains the Swin U-Netr and U-Net to find the optimal coefficient in test time augmentation and then use it as an input of the nnU-Net.} \label{fig2}
\end{center}
\end{figure}

The framework used in this study is shown in Fig.~\ref{fig2} The framework shows that our approach consists of two steps. Learning the appropriate augmentation parameters is the first step. We combined several augmentations with the U-Net and Swin networks to determine how each augmentation affects performance. We devised an optimal combination of test time augmentations using the resulting improvements in network performance when data is augmented.
In the second part of the framework, we get a different pre-trained network - nnU-net - and testing images that need to be segmented. Several augmented images from each testing image were created and then segmented by using the pre-trained nnU-net. These segmentations were combinted into a single segmentation mask using the corresponding augmentation improvement coefficients computed in the first step.

\subsubsection{Augmentation} In the training phase, we used several data augmentation models, including random flip with the spatial axis of 1, 2, and 3, random rotation with the probability of flipping $10\%$, and random shift intensity with the probability of flipping $50\%$ and offset range of $10\%$.

\subsubsection{Optimal augmentation estimation}
One of the ways to improve the existing architectures is to use the test time augmentations~\cite{simonyan2014very} (TTA) method. Since the multimodality dataset is used in this study, the idea of the learnable composition of TTA is proposed. We first trained U-Net~\cite{cciccek20163d} and Swin U-Netr~\cite{cao2021swin} networks to obtain coefficients. During network testing, we modified the testing images with different augmentations to compute how much the segmentation Dice improves using these augmentations. Also, adjust the coefficients of each augmentation so that the transforms with the highest improvement have the largest coefficient. This means that if, for example, the original Dice is $75\%$, and after adding Gaussian noise to the CT and PET data, we reach Dice of $77\%$ and $76\%$, respectively, we increase the CT noise coefficient and reduce the PET noise coefficient.

\subsubsection{validation of optimal augmentations}
To evaluate how well our proposed combination of coefficients works, we needed to test it on a pre-trained model of nnU-Net architecture~\cite{isensee2021nnu}. We augmented the nnU-net with the optimal set of coefficients and tested if it improved the segmentation performance.

\begin{equation}
\frac{1}{n}\Sigma_{i=1}^m \omega_i * A_i = \frac{1}{n} (\omega_1 A_1 + \omega_2 A_2 + \omega_3 A_3 + ... + \omega_m A_m) \ni \Sigma_{i=1}^m \omega_i=n
\end{equation}

In this equation, $A$ is the augmented function(e.g., rotation, add noise, resize, zoom in and zoom out), and $\omega$s are the contribution coefficients obtained in the first stage of the architecture.
We have created an algorithm that determines the best possible combination of contribution coefficients for TTA.

\subsection{Implementation Details}
The configuration of the device used to implement this study is NVIDIA GeForce RTX 3080 GPU with Python 3.8.10 and Torch 1.9.0+cu111. We used patch-based U-Net and Swin U-Netr~\cite{hatamizadeh2022swin} with patch sizes of 96, 96, and 96 in 30000 iterations. Our method was implemented on the nnU-Net~\cite{isensee2021nnu} code and MONAI library~\cite{MONAI_Consortium_MONAI_Medical_Open_2022}. The batch size is 2 for training and 1 for testing. We used an Adam optimizer with a weight decay of 1e-5, and the learning rate is set as 1e-4. In the first step, we separated $12\%$ of the total data for evaluation, $10\%$ for testing, and $78\%$ for learning.

\section{Results}
As we explained earlier, avoiding false positive segmentation in FDG-PET is challenging. For this reason, in addition to considering the Dice coefficient, we also examined the false positive metrics. In training the learnable coefficient phase, U-Net and Swin U-Netr have dice of 0.5039 and 0.5050, respectively. Their false positives are 22.9835 and 22.1624, respectively. As mentioned before, we used a pre-trained nnU-Net to evaluate the proposed coefficients.  For evaluation, we used the five preliminary tests presented in the challenge. In these 5 cases, the augmented nnU-Net has the following performance:

\begin{table}
\centering
\caption{The results of nnU-Net network with the proposed combination of coefficients test time augmentation in the preliminary test set are presented in the AutoPET challenge}
\begin{tabular}{ l| c| c| c}
 Network & Dice Coefficient & False Negative & False Positive\\ 
 \hline
 nnU-Net + TTA & 0.9211 & 1.7865 & 0.9296
\end{tabular}
\label{table:1}
\end{table}

This performance is one percent higher than the performance of the raw nnU-Net, while the false positives decreased from 22.9835 to 0.9296 in contrast to U-Net.

\section{Conclusion}
The proposed method in this study was presented in the automated lesion segmentation in whole-body FDG-PET/CT MICCAI challenge. The network was implemented in this study using an optimal composition of TTA. For this purpose, we trained U-Net and Swin U-Netr to detect the efficient coefficient to improve segmentation performance. Then we used the newly trained U-Net, and Swin U-Netr results with the nnU-Net to prove that our approach improves performance. Averaging the predictions with varying ratios on the augmented data can improve prediction accuracy.

\bibliographystyle{unsrt}
\bibliography{main}
\end{document}